\documentclass[conference]{IEEEtran}
\usepackage[caption=false]{subfig}

\usepackage[]{algorithm2e}

\usepackage{soul}

\usepackage{tcolorbox}
\usepackage{amsmath}
\usepackage{amssymb}
\usepackage[binary-units,per-mode=symbol]{siunitx}
\usepackage{mathtools}
\usepackage{soul}
\usepackage{tikz}
\usepackage[absolute,overlay]{textpos}
\usetikzlibrary{arrows,shadows,decorations.pathmorphing,backgrounds,fit,positioning,calc,shapes}
\usepackage{pgfplots}
\pgfplotsset{compat=newest}
\newlength\figureheight
\newlength\figurewidth

\usepackage{cite}

\begin{document}
\title{Personalized Image-based User Authentication using Wearable Cameras}

\author{
	\IEEEauthorblockN{Ngu Nguyen}
	\IEEEauthorblockA{Aalto University\\
		Email: le.ngu.nguyen@aalto.fi}
	\and
	\IEEEauthorblockN{Stephan Sigg}
	\IEEEauthorblockA{Aalto University\\
		Email: stephan.sigg@aalto.fi}
}

\maketitle

\begin{abstract}

Personal devices (e.g. laptops, tablets, and mobile phones) are conventional in daily life and have the ability to store users' private data.
The security problems related to these appliances have become a primary concern for both users and researchers.
In this paper, we analyse first-person-view videos to develop a personalized user authentication mechanism.
Our proposed algorithm generates provisional image-based passwords which benefit a variety of purposes such as unlocking a mobile device or fallback authentication.
First, representative frames are extracted from the egocentric videos.
Then, they are split into distinguishable segments before a clustering procedure is applied to discard repetitive scenes.
The whole process aims to retain memorable images to form the authentication challenges.
We integrate eye tracking data to select informative sequences of video frames and suggest a blurriness-based method if an eye-facing camera is not available.
To evaluate our system, we perform experiments in different settings including object-interaction activities and traveling contexts.
Even though our mechanism produces variable graphical passwords, the log-in effort for the user is comparable with approaches based on static challenges.
We verified the authentication challenges in the presence of a random and an informed attacker who is familiar with the environment and observed that the time required and the number of attempts are significantly higher than for the legitimate user, making it possible to detect attacks on the authentication system.
\end{abstract}

\section{Introduction}
Personal digital devices are involved in a multitude of aspects of our daily life such as working, entertainment, financial management, navigation, and healthcare.
Consequently, laptops, tablets, and smartphones are versatile appliances that store and operate on private data.
As a protection mechanism, they are usually accessible with secret information known only by the permissible users.
Thus, typical users may have to register and remember at least one password for each device, which is generally related to their personal preferences.
However, the compact size of many of these devices contradicts the deployment of traditional authentication methods.

Graphical passwords, such as the 9-dot log-in screen on the Android platform, have been gaining more popularity than classical text-based approach due to their convenience on the touch-based interface.
Nevertheless, the user is required to remember a static password which is vulnerable to shoulder surfing attacks~\cite{2012_Biddle_GraphicalPasswords}.
Moreover, it is possible to reconstruct the secret pattern from images of the smartphone screen~\cite{2010_Aviv_Smudge}.
Changing the pattern frequently and multi-factor authentication are countermeasures for this attack strategy but they impose more burdens on the users.
Although biometric security mechanisms are increasingly applied for authentication on mobile devices~\cite{2015_Weizhi_SureyBiometric}, they are naturally limited and difficult to substitute in case of being lost or compromised~\cite{2003_OGorman_ComparingAuthentication}.

Wearable cameras allow users to continuously capture videos or images of the environment from the first-person perspective.
They can be mounted to a frame of eyeglasses~\cite{2014_Kassner_PupilDevice}, clipped on a shirt~\footnote{Transcend DrivePro\textsuperscript{TM} Body 10 wearable camera: https://www.transcend-info.com/Products/No-704}, or even attached to a wristband~\cite{2016_Ohnishi_Wrist}.
Modern wearable cameras are equipped with other sensors (e.g. accelerometer and gyroscope) and wireless communication capability (e.g. wifi or bluetooth).
The characteristics of egocentric videos make them a potential candidate for designing authentication schemes on personal devices.
We propose to generate \textit{always-fresh}, \textit{temporal} and \textit{personalized} authentication challenges from images captured by a wearable camera.
A set of photos are displayed on the screen and the legitimate user exploits her implicit knowledge on their temporal order to log-in.
Our main contributions include:
\begin{enumerate}
 \item the first-ever egocentric video-based authentication system with \textit{always-fresh} authentication challenges. The system features eye-fixation and blurriness-based image filters and thereby flexibly supports different camera systems 
 \item the instrumentation and evaluation of two distinct authentication challenge formats in touch-based user interfaces supporting slide-and-swipe gestures: \textit{image-arrangement} and \textit{image-selection}, which achieve the mean entry time of 9.77 seconds and 3.10 seconds, respectively. The measurement is comparable with other memory-based schemes~\cite{2015_Huiping_PassApp}~\cite{2009_Everitt_Passfaces}~\cite{2000_Dhamija_ImageAuthentication} while our mechanism produces \textit{dynamic} graphical passwords (i.e. resistance against shoulder-surfing and smudge attacks)
 \item a method to analyse diverse egocentric videos containing object-interaction scenarios and outdoor scenes. One of our experiment settings spanned three weeks in different European cities
 \item a threat model and an attacking case study with informed, active and random attackers to prove that the adversaries must spend significantly higher effort than the legitimate users on solving the passwords 
\end{enumerate}

\begin{figure}
	\centering
	\includegraphics[width=\columnwidth]{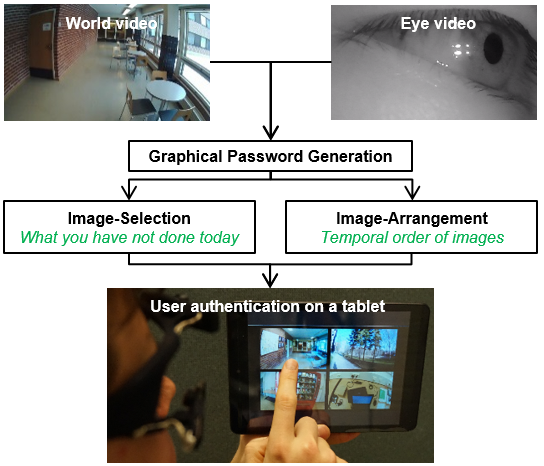}
	\caption{The proposed concept: image-based authentication challenges generated from egocentric videos}
	\label{fig:concept}
\end{figure}

Our proposed authentication scheme is inspired by the \textit{implicit memory} concept, in which previous experiences support the performance of a task without conscious or intentional re-collection of those experiences~\cite{1987_Schacter_ImplicitMemory}.
One popular practice to investigate implicit memory is the priming procedure.
The investigator first gives the participant a set of stimuli (e.g. words, objects, or audio) and then observes the effects in later experiments.
In this paper, the stimuli are video frames recorded by a wearable camera from the first-person perspective.
The images are combined to form the authentication challenge and the user is required to, for instance, (a) rearrange them in the chronological order, or (b) mark images that occurred in a specified time window (e.g. \textit{mark the image(s) that did occur yesterday}).
If our system is equipped with an eye-facing camera, fixations are utilized to detect the user's attention, which further increases the memorability of the chosen image for the user.
Otherwise, we leveraged an image blurriness feature to filter non-informative video frames. 
We employ image segmentation and clustering to select highly memorable images and discard repetitive scenes.
Two authentication formats are implemented and assessed in different scenarios.
The first asks a user to arrange images into their correct timeline while the second requires the users to pick which events have not happened in a certain period of time.
Both formats are graphical passwords and compatible with touch-based interfaces.
Since the authentication challenges are formed from what the users encounter during daily activities, the registration phase is implicit and the passwords are changing over time.
Figure~\ref{fig:concept} illustrates a situation in which the user recalls the chronological order of four images for authentication on a tablet.

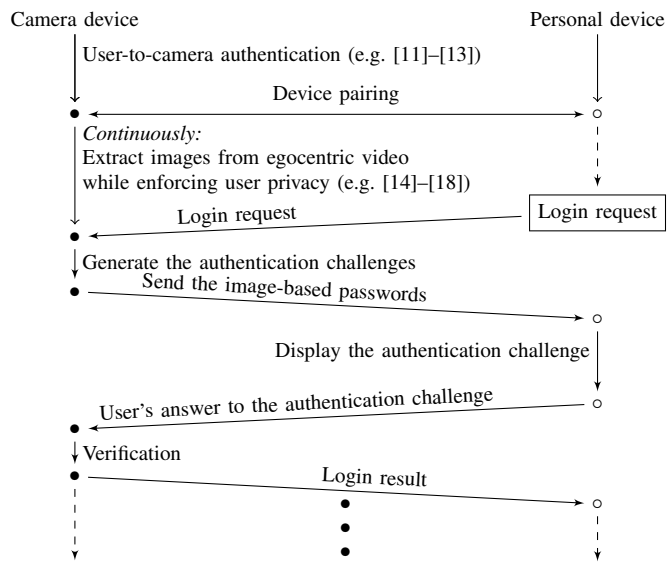
\begin{figure*}
	{\footnotesize
		\begin{tikzpicture}
		\node (A0) at (0,0) {Camera device};
		\node[below=.9cm of A0] (A1) {$\bullet$};
		\node[below=1.3cm of A1] (A2) {$\bullet$};
		\node[below=.4cm of A2] (A3) {$\bullet$};
		\node[below=1.5cm of A3] (A4) {$\bullet$};
		\node[below=.3cm of A4] (A5) {$\bullet$};
		\node[below=0.95cm of A5] (A6) {};
		
		\node[right=5cm of A0] (B0) {Personal device};
		\node[below=.9cm of B0] (B1) {$\circ$};
		\node[below=0.8cm of B1] (B2) {\fbox{Login request}};
		\node[below=0.9cm of B2] (B3) {$\circ$};
		\node[below=.8cm of B3] (B4) {$\circ$};
		\node[below=1.0cm of B4] (B5) {$\circ$};
		\node[below=.6cm of B5] (B6) {};
		
		\node[left=3cm of B5] (C0) {$\bullet$};
		\node[below=-.01cm of C0] (C1) {$\bullet$};
		\node[below=-.01cm of C1] (C2) {$\bullet$};
		
		\setlength{\baselineskip}{12pt}
		\draw[->, align=left]        (A0)--(A1) node [right] {
			
		};
		\draw[->, align=left]        (B0)--(B1) node [right] {
			
		};
		\draw[->, align=left]        (A0)--(A1) node [right] {
			User-to-camera authentication (e.g.~\cite{2015_Hoshen_Biometrics, 2016_Li_HeadMovements,2015_Nigam_SurveyOcular})\\[1.2cm]
		};
		
		\draw[->, align=left]        (A1)--(A2) node [right] {
			\textit{Continuously:}\\[-.1cm]
			Extract images from egocentric video\\[-.1cm]
			while enforcing user privacy (e.g.~\cite{2014_Templeman_PlaceAvoider, 2016_Korayem_Screenavoider, 2014_Rinner_Cartooning, 2016_Wang_SecureSURF, 2011_Denning_ImplicitMemory})\\[1.7cm]
		};
		\begin{scope}[every node/.style={midway},>=latex']
		\draw[<->]        (A1)--(B1) node [sloped,above=1pt] {
			Device pairing
		};
		\draw[dashed, ->, align=right]        (B1)--(B2) node [left] {
			
		};
		\draw[->, align=right]        (B3)--(B4) node [left] {
			Display the authentication challenge\\[-.1cm]
		};
		\draw[<-]        (A2)--(B2) node [sloped,above=5pt,left] {
			Login request
		}; 
		\draw[->, align=left]        (A2)--(A3) node [right] {
			Generate the authentication challenges
		};
		\draw[->]        (A3)--(B3) node [sloped,above=5pt,left] {
			Send the image-based passwords\hspace*{-1.3cm}
		} node [sloped,below] {};
		\draw[->]        (B4)--(A4) node [sloped,above=5pt,left] {
			User's answer to the authentication challenge\hspace*{-2.2cm}
		} node [sloped,below] {};
		\draw[->, align=left]        (A4)--(A5) node [right] {
			Verification
		};
		\draw[dashed, ->, align=right]        (B5)--(B6) node [left] {
		};
		\draw[dashed, ->, align=right]        (A5)--(A6) node [left] {
		};
		\draw[->]        (A5)--(B5) node [sloped,above=5pt,left] {
			Login result\hspace*{-1.3cm}
		} node [sloped,below] {};
		
		\end{scope}
		\end{tikzpicture} 
	}
	\hfill
	\fbox{\begin{minipage}[b]{.9\columnwidth}\small
			\underline{User-to-camera authentication:} 
			The personalized camera device capturing the first person video is a central component in our distributed authentication scheme. 
			We assume that the camera device is at all time strictly associated with the identity of its wearer. 
			To verify identiy of a user from an on-body camera is, for instance, possible using video-based motion analysis~\cite{2015_Hoshen_Biometrics}.
			If an eye tracking camera is available on the headset, user identification based on ocular biometric traits (iris, retina, and eye movement) is possible~\cite{2015_Nigam_SurveyOcular}.
			Also, some smart-glasses are equipped with accelerometers and gyroscopes to facilitate continuous user authentication~\cite{2016_Li_HeadMovements}.
			\\[.2cm]
			
			\underline{Enforcing user privacy on egocentric videos:}
			Another important issue of first-person-view cameras is that they may capture sensitive visual data.
			Recent work has tackled this problem by detecting private locations~\cite{2014_Templeman_PlaceAvoider} and screens~\cite{2016_Korayem_Screenavoider}.
			Recorded video frames can also be transformed to cartoon images~\cite{2014_Rinner_Cartooning} before they are displayed.
			In this case, a legitimate user is still able to recognize the cartoon images if the original ones have been seen before~\cite{2011_Denning_ImplicitMemory}.
			Furthermore, Wang~\textit{et al.}~\cite{2016_Wang_SecureSURF} proposed an outsourcing scheme to protect private images processed on untrusted cloud servers.
		\end{minipage}
	}
	\caption{The protocol for communication during the login procedure include an initial user authentication at the camera device and pairing to the personal equipment. Then, the graphical authentication challenge can be requested and autenticated using the camera device as a verifier for the identity of the wearer.}
	\label{figureAuthenticationScheme}
\end{figure*}

In our image-based authentication mechanism, a wearable camera acts as the password generator or security token without explicitly showing authentication codes.
It empowers user authentication not only on mobile devices (e.g tablets and smartphones) but also on desktop computers, which do not have sensors for such implicit authentication as gait-based method.
Figure~\ref{figureAuthenticationScheme} depicts an authentication procedure that leverages our proposed mechanism.
After the camera device has authenticated the user and paired with the personal device, images are continuously extracted from first person video. 
The image-based authentication challenge is initialized by any log-in request on the personal device. 

In the rest of the paper, we first review related work about recent user authentication schemes and egocentric video summarization methods.
Then, we explain the process of selecting images from first-person-view videos to form the authentication challenges, as well as the techniques to remove non-informative and repetitive scenes.
Finally, our user study results and attacking models are discussed before the paper is concluded.

\section{Related Work}
Pattern-based passwords on touch screens have been comfortably adopted for a long time but they can be compromised, for example, by a camera-based smudge attack~\cite{2010_Aviv_Smudge}.
Thus, several user authentication approaches have been proposed as alternative solutions.
Most of them exploited users' behaviorial data recorded by wearable sensors such as accelerometer~\cite{2016_Li_HeadMovements}~\cite{2016_Sitova_HMOG} or camera~\cite{2016_Song_EyeVeri}.
A couple of innovative authentication forms were built on implicit memory of past information, such as events~\cite{2013_Das_Autobiographical}~\cite{2015_Hang_Questions}, images~\cite{2011_Denning_ImplicitMemory}~\cite{2000_Dhamija_ImageAuthentication}, and installed applications~\cite{2015_Huiping_PassApp}.

The unique characteristics of human movements can serve as an information source for user identification.
Sitov\'{a}~\textit{et al.}~\cite{2016_Sitova_HMOG} introduced a behavior-based authentication mechanism on smartphones.
They record subtle prehensile movements performed by the user when grasping, holding, and tapping to interact with objects.
Specifically, their proposed features are extracted from information captured by accelerometer, gyroscope, and magnetometer during or close to tap events.
The authentication process was implemented and evaluated with scaled Manhattan distance, scaled Euclidean distance, and a single-class Support Vector Machine classifier.
Another behaviroral feature is how human eyes react to stimuli.
Using the built-in front camera on a smartphone, Song~\textit{et al.}~\cite{2016_Song_EyeVeri} introduced a novel smartphone authentication mechanism based on eye movements.
They track human eye movement and extract gaze patterns when the user is exposed to visual stimuli on the phone screen.

The release of smartglasses such as JINS MEME~\cite{2015_MeMe} or Google Glass
has introduced new opportunities and challenges in user authentication.
Li \textit{et al.}~\cite{2016_Li_HeadMovements} takes advantage of head motions when the user wears smartglasses to recognize the wearer.
We usually move our heads in specific patterns when listening to music.
Thus, the authors analysed these distinct movements to identify the wearers.

On the other hand, a pair of smart-glasses equipped with a near-eye screen is able to extend the interface of other devices. 
Winkler \textit{et al.}~\cite{2015_Winkler_GlassUnlock} suggested to employ a numpad layout displayed only on a private near-eye display during the authentication phase.
This approach is a countermeasure against smudge attacks, shoulder-surfing, and camera attacks because the user inputs the password on an almost empty phone interface.
Our proposed approach provides this feature because it shows a new challenge for each log-in attempt.

Implicit memory on images has been used to facilitate user authentication instead of classical character-based passwords.
Dhamija and Perrig~\cite{2000_Dhamija_ImageAuthentication} proposed a novel authentication scheme that relied on the human ability of image recognition.
A user first chooses a portfolio of images from a collection presented by a server.
Then, to be authenticated, the user must distinguish all selected images with decoy ones.
Everitt~\textit{et al.}~\cite{2009_Everitt_Passfaces} studied graphical passwords composed of nine facial images each and the users were authenticated by picking correct faces from distracters.
Denning~\textit{et al.}~\cite{2011_Denning_ImplicitMemory} exposed a user to the complete version of images first and then to their degraded counterparts as the authentication challenge.
Our proposed approach does not require such priming procedure because the user implicitly perceives the scenes, which are captured simultaneously by the wearable camera.
Beyond photos displayed by an external system, Hang \textit{et al.}~\cite{2015_Hang_Questions} proposed to generate the questions for fallback authentication from what users have done with their smartphones.
The information directly comes from the data inside the smartphone.
For example, they asked the user which photo was taken or with whom was communicated in the past.
Indeed, before that, Das~\textit{et al.}~\cite{2013_Das_Autobiographical} had proved the effectiveness of autobiographical authentication through two online questionnaires and one field study.
Applications installed on mobile devices offer a key space for automatic password generation~\cite{2015_Huiping_PassApp}.
Each graphical authentication challenge contains icons of installed apps and decoy apps (collected from the app market).
Whenever the users want to log-in, they make use of memory to recognize the valid icons.
While these approaches~\cite{2013_Das_Autobiographical}~\cite{2015_Hang_Questions}~\cite{2015_Huiping_PassApp} rely on data stored on the mobile devices, ours exploits visual information recorded by a wearable camera during a user's daily life.

Recently, the proliferation of wearable cameras has opened up an emerging research trend: the analysis of egocentric videos~\cite{2015_Betancourt_Survey}, in which computer vision and image processing techniques have been utilized to analyse the first-person-view visual data.
The optical flow extracted from these videos depicts the movements of human head and body, from which the user's activities can be inferred~\cite{2014_Poleg_Segmentation}.
Those movements are distinctive enough to facilitate wearer identification~\cite{2015_Hoshen_Biometrics}.
To cope with the vast amount of these images and videos, summarization algorithms are leveraged to retain meaningful and memorable excerpts~\cite{2013_Lu_Summarization}~\cite{2015_Xu_GazeSummarization}.
The outcomes are context-dependent multimedia contents that are individualized to a certain user.
Technically speaking, our work is similar to video summarization~\cite{2013_Lu_Summarization}~\cite{2015_Xu_GazeSummarization} or snap point detection~\cite{2014_Xiong_Snap} in terms of extracting important parts of the egocentric videos.
In a seminal work, Lu \textit{et al.}~\cite{2013_Lu_Summarization} relied on objects and people that the user interacts with to compute an importance metric.
Then, they leveraged a story-driven approach to organize the summary of egocentric videos.
For important metric computation, it is possible to use a regressor operating on the distance between objects and hands, distance between objects and the center of the frame (as an estimation of the gaze), etc.
The story-driven approach was also utilized by Xiong \textit{et al.}~\cite{2015_Xiong_Storyline} to extract a storyline representation for first-person videos.
Their system even supports story-based queries based on four elements including \textit{actors}, \textit{locations}, \textit{objects}, and \textit{events}.
Using a Markov Random Field, they model the hierarchical dependency between those aforementioned elements.
Because gaze is a natural measurement of how much the user is attracted to a video subshot, it is used by Xu \textit{et al.}~\cite{2015_Xu_GazeSummarization} in egocentric video summarization.
The authors relied on gaze information to segment videos before merging them into subshots.
Region-based Convolutional Neural Networks were used to extract the feature vector in the central frame of each subshot.
Another approach is based on photos captured intentionally and uploaded online to select important frames in egocentric videos~\cite{2014_Xiong_Snap}.

Visual context recognition is a further area closely related to our research.
Furnari \textit{et al.}~\cite{2015_Furnari_Contexts} analysed egocentric videos to facilitate the recognition of personal contexts.
They recorded the videos with different camera models in various locations.
Each image is represented with a holistic descriptor (or the ``gist'' of the scene), local features (Scale-invariant feature transform or SIFT) encoded by Improved Fisher Vectors, and values from the penultimate layer of Convolutional Neural Networks.
Castro \textit{et al.}~\cite{2015_Castro_EgocentricActivities} combined Convolutional Neural Networks and Random Decision Forests to identify human activities from egocentric images.
Even though our work focuses on personal surroundings, we do not define a fixed set of contexts to be identified.
In particular, our image selection method, which relies on segmentation and clustering, aims to group video frames based on their visual characteristics without prior knowledge on the contexts and contents.

The fundamental components of our authentication scheme is built on image analysis techniques.
We first extract the visual descriptors from each frame of the egocentric videos and then select representative frames that the user can easily memorize.
In order to do that, the selected images should contain particular and salient details which help the user to remember their order of appearance.
%This scheme therefore relies on the user's \textit{implicit memory} about the daily routine.

\section{Image-based Password Generation}
\label{sec:algorithm}

In this section, we discuss the technical details behind our graphical password generation process.
First, several representations of authentication challenges are recommended, along with their use cases.
Second, we present our algorithm to select candidate video frames which are later used to create image-based authentication challenges.
The general idea is that the content in those frames triggers the user's \textit{implicit memory}~\cite{1987_Schacter_ImplicitMemory}.
Finally, we describe the features extracted from the frames of egocentric videos.

\subsection{Graphical Password Design}
\begin{figure}
	\centering
	\includegraphics[width=\columnwidth]{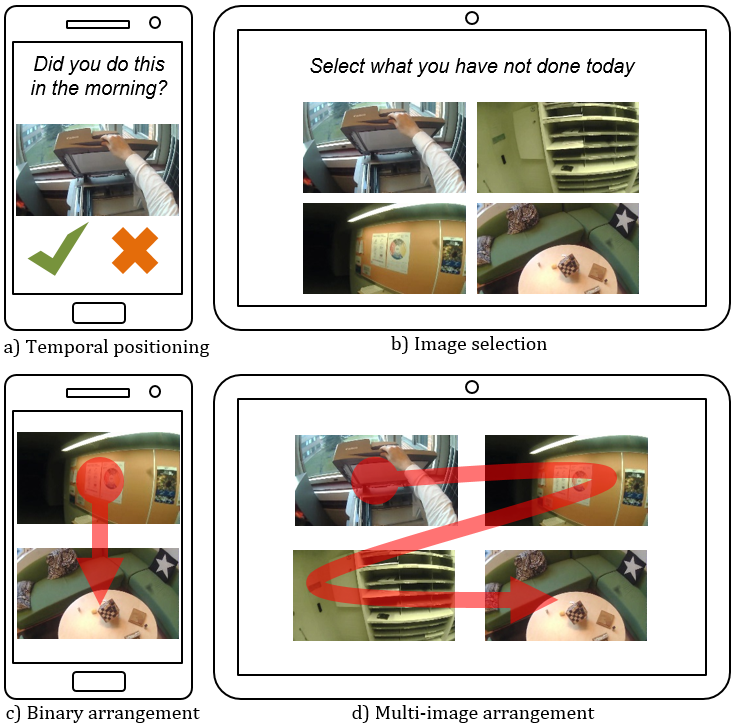}
	\caption{Proposed screen layouts of our authentication mechanism}
	\label{fig:interface}
\end{figure}

Our framework described in Section~\ref{sec:algorithm} takes egocentric videos and exports images, which are split into different contexts along the timeline.
These contexts are related to the wearer's activities or locations.
We suggest four authentication schemes where the output images can be employed as shown in Figure~\ref{fig:interface} (A sequence of challenges may appear consecutively to harden the challenge):
\begin{itemize}
	\item Temporal positioning (Figure~\ref{fig:interface}a): We leverage the moment when the images are captured to produce such \textit{Yes/No} questions as \textit{``Did you do that ...?''}.
	\item Image selection (Figure~\ref{fig:interface}b): The user has to select which events (scenes) have not been done (or has been done yesterday, or similar). 
	\item Binary arrangement (Figure~\ref{fig:interface}c): The system introduces two images and the user slides to determine the right chronological order. %The probability of a successful guess is $\frac{1}{2}$.
	\item Multi-image arrangement (Figure~\ref{fig:interface}d): Multiple images are shown on the screen randomly and a user arranges them in the right temporal order. Our implementation changes the images whenever a wrong answer is submitted, which makes guessing attacks more challenging.
\end{itemize}
The aforesaid options are suitable for various types of authentication purposes.
The first three schemes can be used as instant log-in (one challenge only, depending on the user settings).
The last one may require longer duration of consideration, which makes it appropriate for fall-back authentication.
In our experiments, the \textit{image-arrangement} and \textit{image-selection} authentication scheme are implemented and evaluated (others can be extended from these).

\subsection{Selecting Representative Frames}

\begin{figure}
	\centering
	\includegraphics[width=70mm]{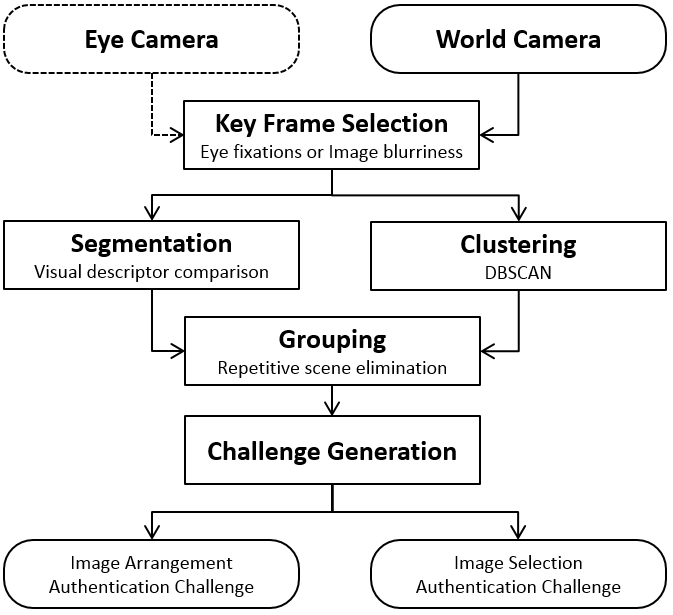}
	\caption{The process of genarating authentication challenges from first-person-view videos}
	\label{fig:selection}
\end{figure}

We propose to choose candidate images for forming the authentication challenges based on two criteria:
\begin{itemize}
	\item Memorability: The image must be as easy to remember for the wearer as possible.
	\item Popularity: The image should not be so common that it can be easily recognized by the adversary. 
	This can be considered as the uniqueness measure of an image frame.
\end{itemize}

We aim to select frames that contain clear information (i.e. there are details that can help the user to infer the chronological order).
Figure~\ref{fig:selection} details the procedure, which consists of several image analysis techniques, including: key frame selection, video segmentation, and scene clustering.
After these processing steps, we form the authentication challenges of two formats: \textit{image-arrangement} and \textit{image-selection}.

\textbf{Key Frame Selection}:
To reduce the number of images extracted directly from the recorded videos, we retain only key frames.
In order to do that, there are two options: (i) selecting frames whenever a fixation appears and (ii) retaining images at every specific moment (e.g. five images per second).
The first approach requires data from the eye tracker device, which may not be universally available in all wearable cameras.
The second requires a filtering method and more computational demand.
Eye fixations provide essential information of the wearer's visual behavior and have proven useful in video summarization~\cite{2015_Xu_GazeSummarization} because they reveal the user's attention.
For example, there is a higher probability that an image is important and memorable if the wearer's eyes focuses while the image is captured.
For the second solution, we aim to remove unclear video frames caused by body and head movements.
To do that, we calculate the blur information of each frame with the method of Crete~\textit{et al.}~\cite{2007_Crete_Blur}.
Formally, suppose the video consists of $k$ frames $v_i$ whose the blurriness is $n_i$, we select $v_i$ if $n_i \geqslant median(\{n_i | i \in \mathbb{N}, 1\leqslant i \leqslant k\})$.
We observed that image sequences produced by two methods were similar and equally useful.

\begin{figure}
	\centering
	\includegraphics[width=\columnwidth]{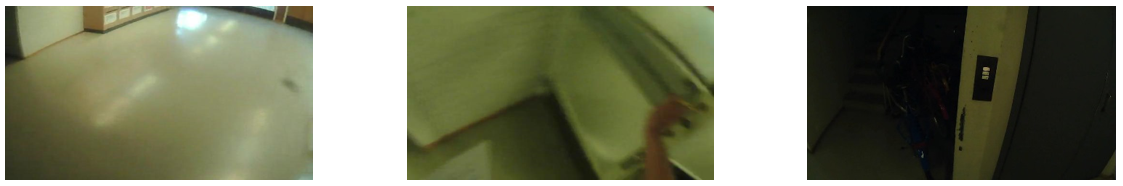}
	\caption{Non-informative images discarded from small clusters}
	\label{fig:filtered}
\end{figure}

\begin{figure*}
	\tcbset{width=(\linewidth-2mm)/2,nobeforeafter,arc=1mm,
		colframe=black!75!white,colback=white,fonttitle=\bfseries,fontupper=\small,
		left=2mm,right=2mm,top=1mm,bottom=1mm,equal height group=parbox}
	\begin{tcolorbox}[parbox,adjusted title={Image-arrangement}]
		\textit{Image-arrangement} passwords require users to recall the relative timeline order of images.
		Thus, the input data is treated as a continuous sequence of videos.
		After clustering, we group video frames in such a way that images in the same segment belong to the same cluster.
		Then, frames which are in the same cluster but appear in different segments (i.e. repetitive scenes) are removed from the candidate image set.
		This ensures that there is no pair of images describing the same action but is interleaved in time domain by images belonging do a different action, since this can cause confusion for the user to guess the occurrence order.
		Thus, our algorithm reduces the popularity of repetitive scenes while strengthening the memorability.
	\end{tcolorbox}\hfill%
	\begin{tcolorbox}[parbox,adjusted title={Image-selection}]
		\textit{Image-selection} passwords insist on the moment when the user executes a certain activity or observes a particular scene.
		In this paper, the authentication question \textit{``What have you not done today?''} is showed above the images.
		Hence, we supply the clustering algorithm with videos from two consecutive days.
		The results are processed in such a way that scenes appear in both days are discarded.
		Note that in this case we do not remove repetitive segments happening in the same day because they do not influence the user's selection.
		This type of passwords can be extended to other temporal scales.
		For example, the question \textit{``What have you not done after 11:00am?''} separates events based on the recording time.
	\end{tcolorbox}%
	\caption{Details on two authentication challenge formats}
	\label{fig:details}
\end{figure*}

\begin{figure*}
	\centering
	\includegraphics[width=180mm]{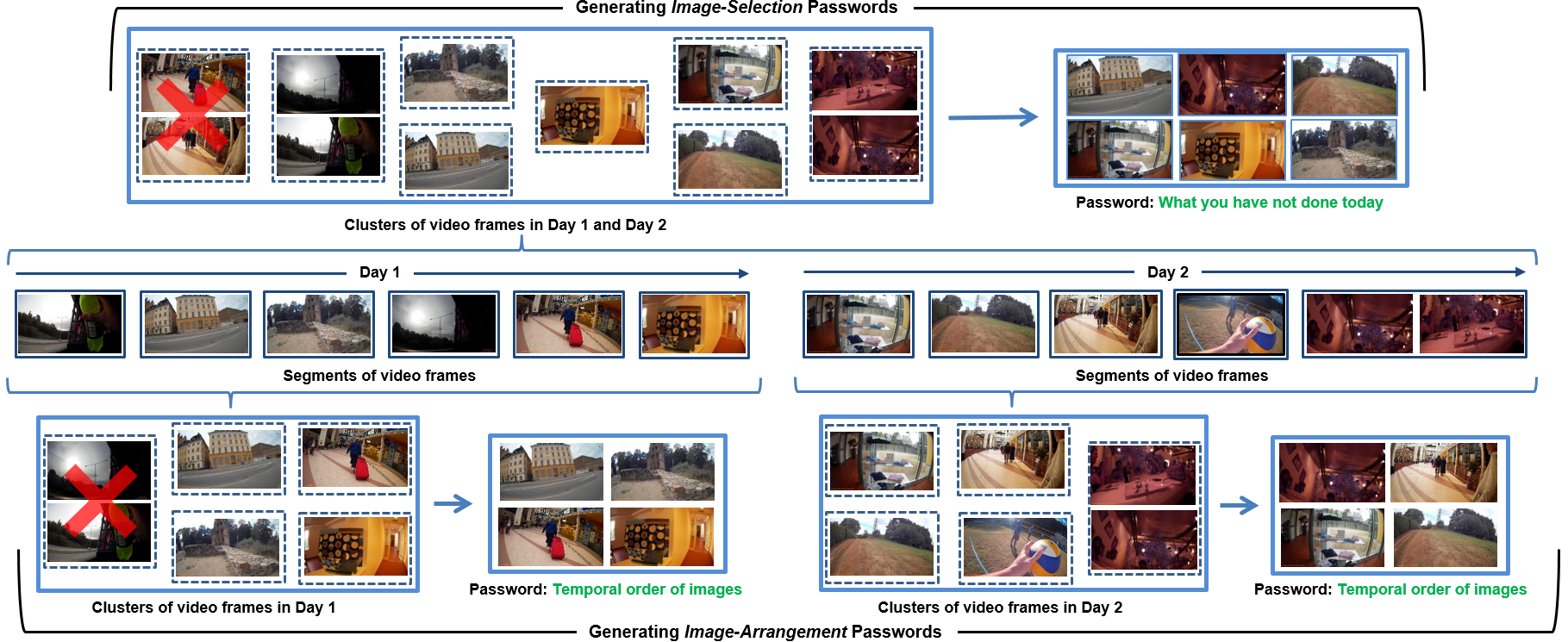}
	\caption{Using image segmentation and clustering to remove repetitive scenes and form the authentication challenges}
	\label{fig:segmentcluster}
\end{figure*}

\textbf{Segmentation}:
After the first phase of extracting key frames, the $k^\prime \leqslant k$ selected images are then segmented into scenes, based on similarity of visual features, which are described in Section~\ref{sec:features}.
That means if the difference between two frames is below a certain threshold, they belong to the same segments.
In this paper, we define the threshold as the median value of the Euclidean distance between two consecutive (representative) frames in the feature space.
Suppose $f$ is the function that extracts the feature vector from each video frame, we define the threshold as $median(\{ d(f(v_i), f(v_{i+1})) |  i \in \mathbb{N}, 1\leqslant i \leqslant (k^\prime - 1)\})$ where $d$ is the Euclidean distance function.
We choose this threshold instead of the pairwise distances between every pair of frames for its lower computational complexity ($\mathcal{O}(n)$ instead of $\mathcal{O}(n^2)$). 
The preliminary analysis of our experiments proved that it produced better segmentation results.
This parameter, as well as the aforementioned blurriness limit, is adapted to the personalized videos.
We experimented with cosine similarity but it did not improve the result in our case.
In addition, the user is able to specify the timeline in which videos are analysed to form the authentication challenges.
Thus, even within a specific user's clip collection, the values of these thresholds depends only on the videos themselves.

\textbf{Clustering}:
The next step is to cluster similar segments into groups of similar scenes.
The number of clusters is not determined beforehand because there is no knowledge on which activities are repetitive.
To fulfil this requirement, we use \textit{Density-based spatial clustering of applications with noise} (DBSCAN)~\cite{1996_Ester_DBSCAN}, which groups together points that are in a neighborhood.
The algorithm requires two parameters: the distance threshold to group similar images and the minimum number of images in each cluster.
The similarity threshold of images is determined by analysing the difference of consecutive frames.
We use the same threshold value for both segmentation and clustering.
The second parameter allows us to discard \textit{noise} images, which pop up abruptly and are \textit{non-informative}.
Some example video frames that are removed by this method are shown in Figure~\ref{fig:filtered}.
They are blurred, dark or do not contain interesting objects (almost empty).

\textbf{Grouping \& Challenge Generation}:
The challenge is generated from the respective clusters as depticed in figure~\ref{fig:segmentcluster}. 
We have considered two distinct authentication challenge formats as detailed in figure~\ref{fig:details}.
To generate an \textit{image-selection} authentication challenge, scenes occurring in both days are eliminated while repetitive activities in the same day still appear in the final password.
The lower part of the figure sketches one \textit{image-arrangement} password for each day, in which similar images belonging to different segments are grouped into the same cluster.
Such photos are not selected to form the authentication challenge.
Note that the selected images belong to different contexts (e.g. on bus, at the train station, hiking, playing volleyball, having dinner, etc).
When generating the graphical passwords, we select one representative frame from one scene (i.e. segment) to guarantee the distinguishability of the images.

\subsection{Image Features}
\label{sec:features}

\begin{figure}
	\centering
	\includegraphics[width=\columnwidth]{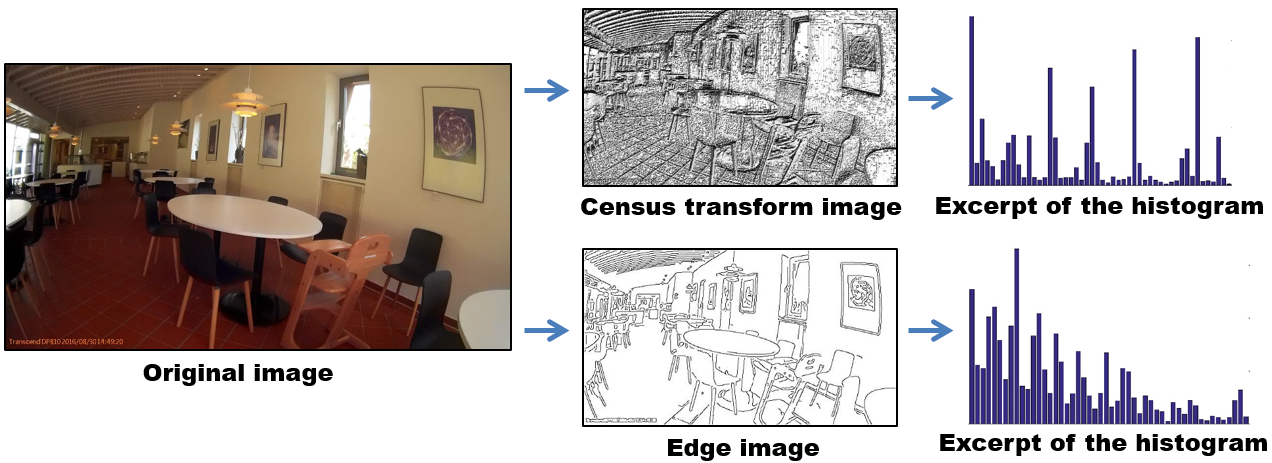}
	\caption{Features extracted from a video frame (We only show excerpts of the histogram-based image descriptors for better visulization)}
	\label{fig:features}
\end{figure}

\begin{figure*}[htp]
	\centering
	\includegraphics[width=180mm]{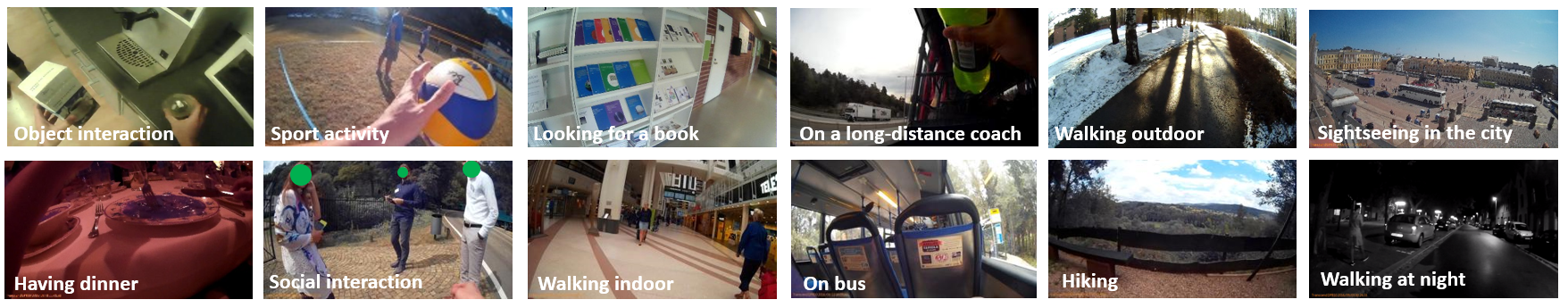}
	\caption{Sample images extracted from our egocentric videos. They vary in activities (e.g. holding objects, playing sport, traveling on public transport, and social interaction), locations (indoor and outdoor), weather conditions (e.g. winter and summer), and the light intensity (e.g. sunny day, having dinner at night, and dark scene).}
	\label{fig:dataset1}
\end{figure*}

\begin{figure}[htp]
	\centering
	\subfloat[Head-mounted camera]{\includegraphics[height=40mm]{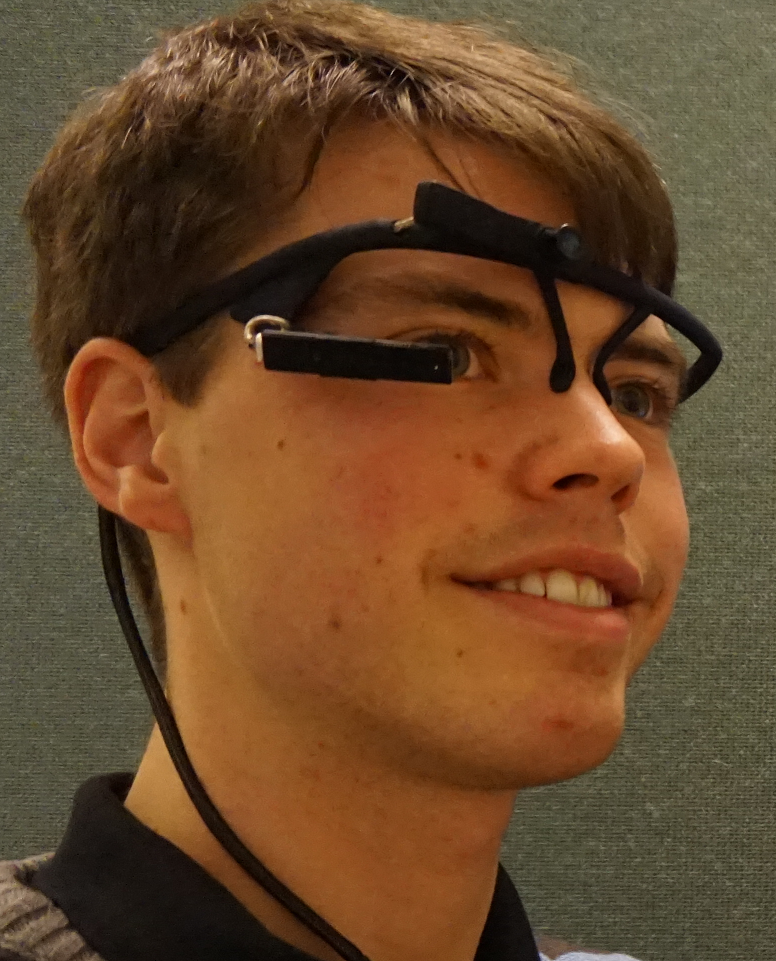}	\label{fig:head}}\hfill
	\subfloat[Chest-mounted camera]{\includegraphics[height=40mm]{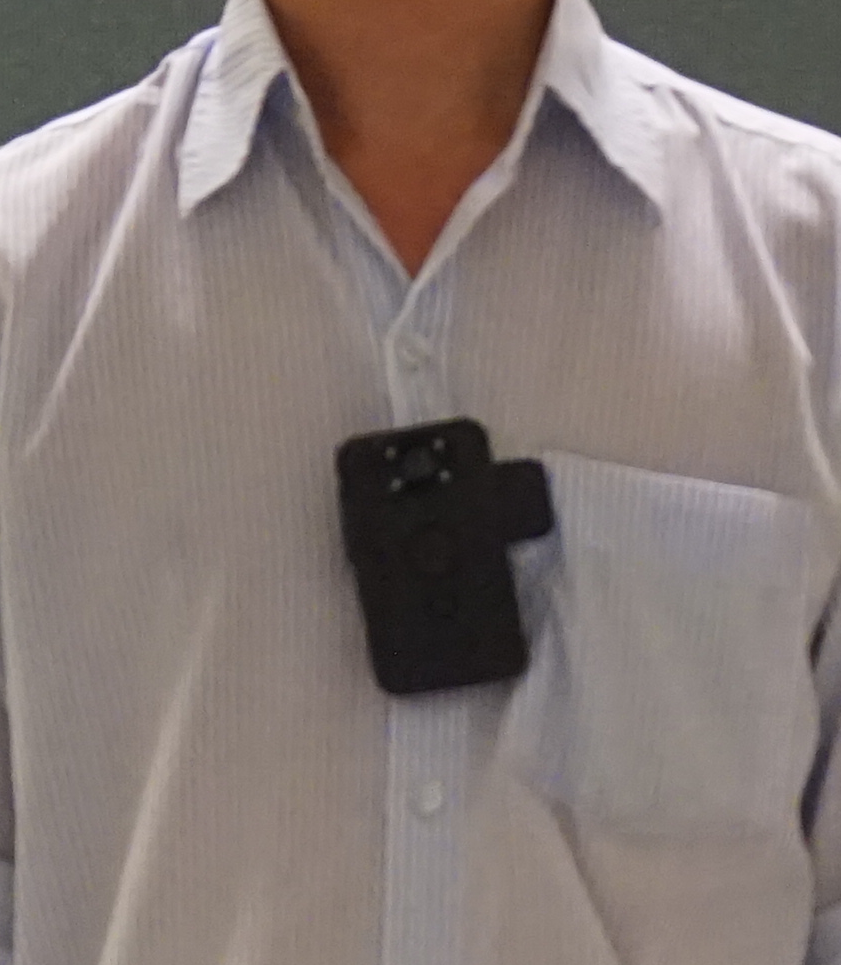}		\label{fig:chest}}
	\caption{The devices used to collect video data}
	\label{fig:device}
\end{figure}

We aim to extract both global and local characteristics of each video frame.
The feature vector must therefore be able to describe the appearance of the whole scene as well as the local sub-regions and individual objects.
To fulfil this requirement, we combine Census Transform Histogram (CENTRIST) and the Pyramid of histograms of Orientation Gradients (PHOG) to form the representation of our first-person-view images.
We perform Principle Component Analysis to reduce the number of dimensions.
In the experiments, we achieved the best result when each feature vector contains $n = 100$ variables.
The final feature vectors are then used to split the video frames into temporal sequences and cluster images that have similar content.
Figure~\ref{fig:features} shows an example image, its intermediate representations, and excerpts of its visual descriptors.

\textbf{CENTRIST}:~\cite{2011_Wu_CENTRIST} generates a holistic representation of a scene by capturing the structural characteristics and the rough geometry.
Census transform compares the intensity value of each pixel with those of its neighborhood.
If the value is less than one of its neighbors, the corresponding location is assigned a bit 0, otherwise, a bit 1 is used.
Then, eight bits in the neighborhood are concatenated and converted to a base-10 number called census transform value for the central pixel.
After all pixels of the image are evaluated, the CENTRIST descriptor is constructed from a histogram of census transform values.
A spatial pyramid scheme is applied to obtain a robust global representation.
An image is split into (sliding) blocks in different levels.
In each block, the CENTRIST descriptor is extracted independently.
Finally, all descriptors (i.e. histograms) of all blocks at all levels are aggregated to form the representation of the whole image.

\textbf{PHOG}:
To complement the above global descriptors, we compute the PHOG feature~\cite{2007_Bosch_PHOG}.
With this descriptor, Bosch~\textit{et al.}~\cite{2007_Bosch_PHOG} aimed to represent an image through its local shape and the spatial relations of the shape.
They divide an image into rectangular regions at several resolutions and calculate the distribution (histogram) of edge orientations in each region.
Each bin of the histogram contains the number of edges whose orientations belong to a specified angular range.
Then, the concatenation of histograms from all regions becomes the descriptor of the image.

\section{User Study and Discussion}
In this section, we present our case studies to verify the feasibility of the approach.
The videos were collected by wearable cameras, then transmitted to a computer for frame extraction and password generation.
After that, the wearer solved the passwords on other devices while the number of attempts and the entry time were collected for further analysis.
We assume that all involved components (e.g. wearable cameras, personal devices, processing servers, etc) are securely connected or paired in advance, using wired or wireless communications.

We performed experiments in three scenarios under various conditions:
\begin{itemize}
	\item a pilot study with two wearable cameras to familiarize the users (see Figure~\ref{fig:device} for the mounting positions)
	\item an object-interaction setting in an office
	\item a daily condition study in which subjects wore the camera during several days
\end{itemize}
Some images extracted from our video clips are shown in Figure~\ref{fig:dataset1}.
Those photos contain diverse details due to the subjects' activities, body movements, locations, weather, light intensity, and scene appearance.

\subsection{Pilot Study}
\label{sec:pilot}

This study aims to introduce our authentication concept and verify its feasibility.
The \textit{image-arrangement} scheme is leveraged in these experimental sessions.

We employed the headset produced by Pupil Labs~\cite{2014_Kassner_PupilDevice} to acquire the world and eye videos.
This device comprises two cameras on a frame: the \textit{world} camera to capture the scene from the user's point-of-view and the \textit{eye} camera to record videos of one user's eye.
Along with the hardware, a software package is provided with such functions as: camera control, eye tracking and fixation detection, visualization, etc.
We attached a laptop to the headset through a USB connection in order to store both video streams simultaneously since this is currently not supported with the mobile app.
Figure~\ref{fig:head} shows our device installation for the data collection phase.
The video resolution is reduced from $1280\times720$ to $320\times180$.

An application was developed with Matlab~\footnote{Matlab: https://www.mathworks.com/} in order to display the authentication challenges.
Each challenge, i.e. graphical password, consists of four video frames which are selected from distiguishable temporal segments (for example, see~Figure~\ref{fig:segmentcluster}).
The users clicks on each frame to form a chronological sequence of images.
Each image stems from one temporal segment and contains visually-diverse information.
Due to the clustering procedure, we do not select repetitive actions, which may cause confusion for the user when selecting the right chronological order.
In each challenge, the user needs to answer $n = 4$ image-based authentication challenges consecutively.
All passwords are generated from the same video source but the images are selected randomly from different segments.
The number of attempts and the entry time are collected for each password.

We recruited five subjects (two females, two wearing glasses) to participate in two sessions.
The first one aimed to make the users familiar with the system while the second encouraged the users to discover the environment and perform more activities.
The duration of videos in the former is $3 - 5$ minutes and $6 - 12$ minutes in the latter.
The subjects in our experiments wore the device and went to different locations.
We suggested possible places to visit but the subjects did not need to follow any specific routine.
After that, the system asked them to organize four images into the right chronological order.
The number of attempts and the entry time are recorded for further analysis.
Since the entry time includes the unsuccessful log-in trials, we measure the time until a subject solves a password successfully regardless the number of attempts, which is equivalent to the \textit{total time} concept proposed by Everitt~\textit{et al.}~\cite{2009_Everitt_Passfaces}.
We analysed the users' effort in the second session because the first one only introduced our system to the subjects.
The average number of attempts and entry time were 1.65 times and 19.07 seconds, respectively.
Particularly, in our experiments, the fastest time that a user solved an authentication challenge is 3.79 seconds.
These results indicate that our approach can be utilised to generate the visual authentication challenges from egocentric videos.

Next, we aim to improve the usability because the Matlab-based prototype only supports mouse-click interaction.
Thus, we developed a web application that supported slide-and-swipe gestures with Javascript and HTML.
To record the videos, a Transcend DrivePro\textsuperscript{TM} Body 10 device was worn by each subject in two consecutive days (cf. Figure~\ref{fig:chest}).
The camera has a rotational clip to attach on clothes or backpack straps.
Whenever possible, the camera wearer started to continuously record videos with the consideration of technical, legal, and social regulations.
The videos contained visual data of the subjects' daily activities.
The system generates a new authentication challenge whenever the user wants to log-in.
It changes the challenge (i.e. new images) if the user submits a wrong arrangement.
We conducted the same evaluation process and achieved better results.
Specifically, the mean entry time is 9.79 seconds and the mean number of attempts is 1.87.
Table~\ref{tab:pilot} summarizes the user effort in the Matlab-based and web-based interface for comparison.
There is an observation that the web-based interface allows more attempts but less time-consuming.
Due to more variation in the two-day video clips, the user effort may be higher but the touch-based interface decreases the entry time significantly.
Thus, we customized this kind of user interface for other experiments.

\begin{table}[]
	\centering
	\begin{tabular}{llll}
		\cline{1-3}
		\multicolumn{1}{|l|}{\textbf{Application}}          & \multicolumn{1}{c|}{\textbf{Matlab-based}} & \multicolumn{1}{c|}{\textbf{Web-based}} &  \\ \cline{1-3}
		\multicolumn{1}{|l|}{\textbf{Entry time (seconds)}} & \multicolumn{1}{l|}{19.07 (18.17)}         & \multicolumn{1}{l|}{9.79 (7.00)}        &  \\ \cline{1-3}
		\multicolumn{1}{|l|}{\textbf{Number of attempts}}   & \multicolumn{1}{l|}{1.65 (1.09)}           & \multicolumn{1}{l|}{1.87 (1.37)}        &  \\ \cline{1-3}
		&                                            &                                         & 
	\end{tabular}
	\caption{User effort on the Matlab-based (mouse-click interaction) and web-based (slide-and-swipe interaction) interface in our pilot study. The numbers in brackets are standard deviation.}
	\label{tab:pilot}
\end{table}

\subsection{Object-interaction Scenario}
\label{sec:room}

\begin{figure*}[htp]
	\centering
	\subfloat[Average number of attempts to solve each password]{\includegraphics[width=\columnwidth]{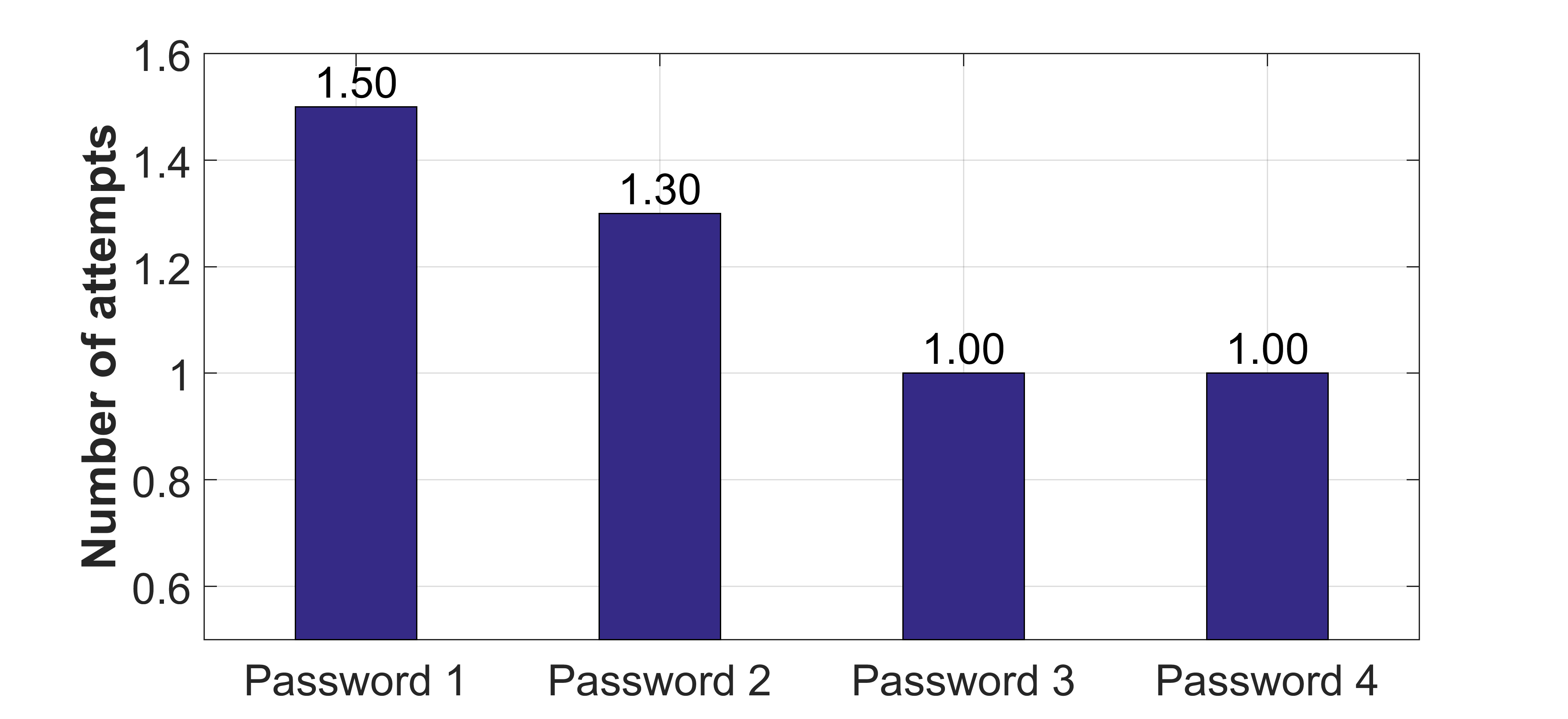}	\label{fig:attempts_room}}
	\hfill
	\subfloat[Average entry time spent on solving each graphical password]{\includegraphics[width=\columnwidth]{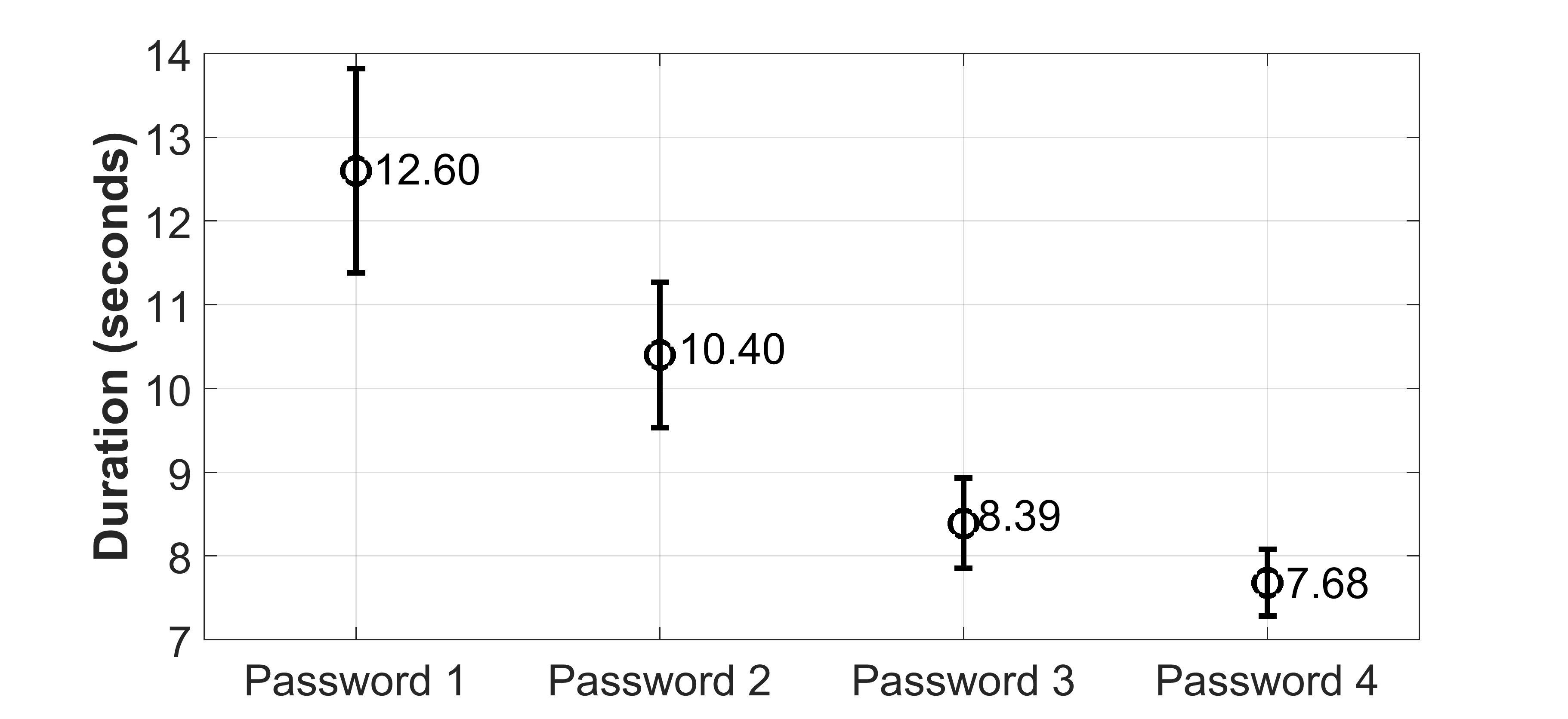}	\label{fig:time_room}}
	\caption{User effort in terms of the number of attempts and the entry time in the object-interaction scenario with \textit{image-arrangement} authentication challenges. The prototype supports slide-and-swipe interaction to decrease the entry time.}
	\label{fig:room}
%	\hspace*{\fill}%
\end{figure*}

We prepared a room for object-interaction activities.
The setting was challenging to our system because all activities occurred at the same location.
Thus, the object appearance and interaction are the only visual cues that assist our video analysis algorithm.
The room in this experiment was a usual workplace in which there are basic furniture and office equipment.
The room is new to all seven participants (four females, two with glasses) in the experiment.
Each subject was asked to perform at least five activities (possible activities were suggested but the subjects were free to perform also others and on the order), then solve passwords generated from the recorded egocentric videos.
The suggested activities include: reading a book, working on a laptop, writing on paper, writing on a board, viewing a poster, talking to a person, using a smartphone, unboxing an item, playing a board game, and using a paper-cutter.
The objects are put on the desks (laptop, paper, gameboard, and paper-cutter), hung on the wall (poster and board), or in the subject's pocket (smartphone).
The person who communicates with the subject is in the same room.

In this scenerio, we used the Pupil headset as in the previous study.
The \textit{image-arrangement} scheme was investigated in this setting.
We built a web-based interface to simulate a touch screen because it decreased the entry time and improved the usability (see Section~\ref{sec:pilot}).
The web application could be accessed from desktop computers or mobile devices.

To assess the users' effort, we asked each subject to answer a sequence of different image-based passwords.
To make the authentication scheme more secure, a new password is generated whenever the user forms a wrong sequence of images in the chronological order.
The statistical number of attempts and entry time are showed in Figure~\ref{fig:room} for the first four passwords.
There is a declining trend in both Figure~\ref{fig:attempts_room} and Figure~\ref{fig:time_room} indicates that the user can learn about the occurrence order of images after solving a couple of passwords.
Our experimental participants spent in average 12.62 seconds to answer the first authentication challenge.
When all four passwords were taken into account, each subject spent 9.77 seconds with 1.21 attempts to solve a password.
The duration is comparable with other graphical password schemes such as PassApp~\cite{2015_Huiping_PassApp} 7.27 seconds, Passfaces~\cite{2009_Everitt_Passfaces} 18.25 seconds, and D\'{e}j\`{a} Vu~\cite{2000_Dhamija_ImageAuthentication} 27 - 32 seconds.
In addition, our proposed scheme exposes a new password if the previous attempt results in a wrong temporal sequence of images.

\subsection{Daily Condition Study}
\label{sec:daily}

\begin{figure}
	\centering
	\subfloat[Number of clicks to solve the passwords]{\includegraphics[width=\columnwidth]{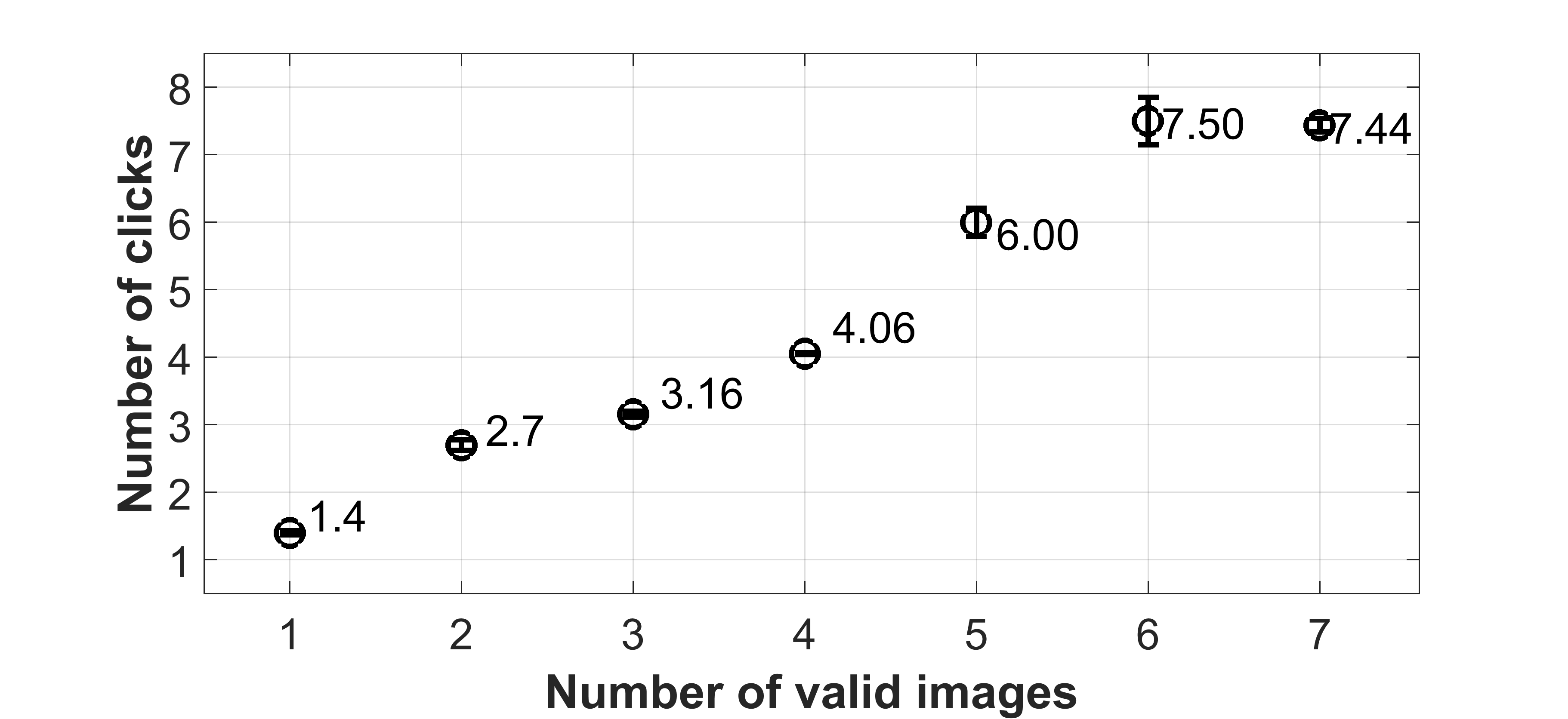}	\label{fig:clicks_long}}\hfill
	\subfloat[Time duration spent on solving the passwords]{\includegraphics[width=\columnwidth]{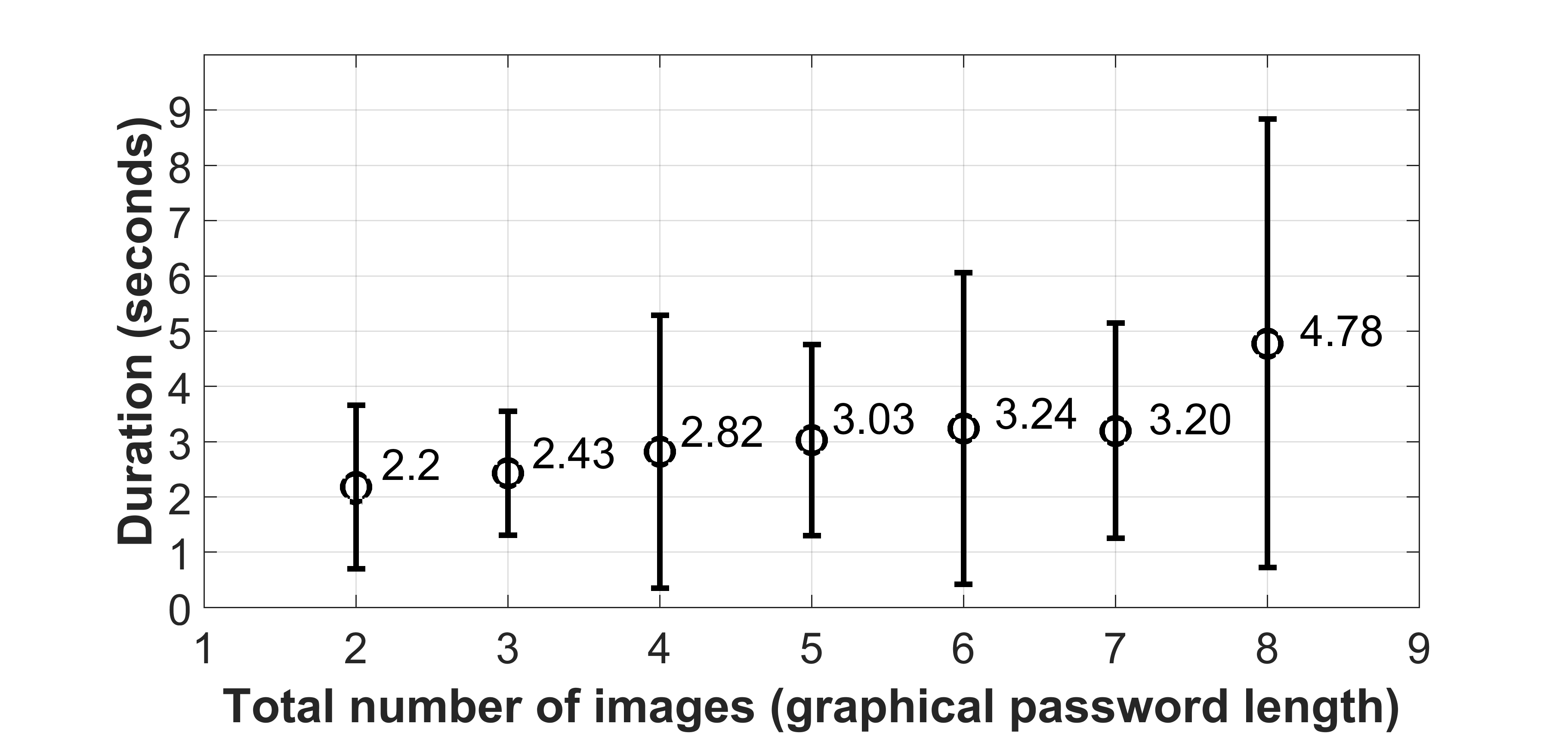}	\label{fig:time_long}}
	\caption{User effort in terms of the number of clicks and the entry time (duration) in the daily condition study with \textit{image-selection} authentication challenges}
	\label{fig:long}
	\hspace*{\fill}%
\end{figure}

To evaluate our system in the long-term context, we analysed two scenarios in which the subjects wore the cameras for several days and answered the authentication challenges.
In this study, we focus on the \textit{image-selection} authentication scheme.
A similar experiment had been deployed with the \textit{image-arrangement} scheme in the latter part of Section~\ref{sec:pilot}. 

In the first case, the Transcend DrivePro\textsuperscript{TM} Body 10 camera were worn by a graduate student in three weeks (cf. Figure~\ref{fig:chest}).
Mainly, the videos contain scenes of home, workplace, and navigation (both on foot and on public transport).
The subject even captured videos of a trip to Stockholm (Sweden), Wadern (Germany), and Brussels (Belgium), where the student visited for the first time.
The subject performed the log-in actions multiple times during the experiment with dynamically-generated graphical passwords.

Another web-based prototype was implemented that displays from two to eight photos.
They came from videos recorded in two consecutive days.
The subject was required to pick the valid images describing events in the previous day but not in the current day.
Our system randomized both the total number of images $n$ and the quantity of valid photos $k$ and did not let the user know those parameters.
In a password containing $n$ images (i.e. password lengh is $n$), there must be $1 \leq k \leq (n - 1)$ valid ones.
The user may select and unselect an image multiple time until achieving the correct configuration.
In the web application, clicking on an image alternates its state.
We recorded the number of clicks instead of the number of attempts.
The entry time is calculated from when the password fully appears until when the proper selection is reached.

Regardless of password length and number of valid images, the user consumes 3.10 seconds to click 5.14 times on the images on average.
Figure~\ref{fig:long} reveals the relationship between the password configuration, i.e. length and number of valid images ($n, k$), and the user's effort.
Apparently, the number of clicks and the entry time increase when the password becomes complex.
In the experimental results, 50\% of the two-image passwords were solved in about 1 second or less with a single click.
It makes this password length a suitable candidate for quick unlocking a mobile device.
In order to investigate the most secure configuration (i.e. eight-image passwords), we adjust the system in such a way that it yields longer authentication challenges (approximately 50\% of passwords contain eight photos).
Even though these are challenging, the user solved them on average 4.78 seconds (cf. Figure~\ref{fig:clicks_long}) with 7.44 clicks (cf. Figure~\ref{fig:time_long}).

\begin{table}[]
	\centering
	\begin{tabular}{lllll}
		\cline{1-4}
		\multicolumn{1}{|l|}{\textbf{Correctness}}          & \multicolumn{1}{c|}{\textbf{$\geqslant$ 50\%}} & \multicolumn{1}{c|}{\textbf{$\geqslant$ 75\%}} & \multicolumn{1}{c|}{\textbf{100\%}} &  \\ \cline{1-4}
		\multicolumn{1}{|l|}{\textbf{Entry time (seconds)}} & \multicolumn{1}{l|}{3.77 (2.88)}                 & \multicolumn{1}{l|}{4.85 (3.08)}                 & \multicolumn{1}{l|}{5.66 (3.33)}    &  \\ \cline{1-4}
		\multicolumn{1}{|l|}{\textbf{\# clicks}}   & \multicolumn{1}{l|}{2.09 (0.98)}                 & \multicolumn{1}{l|}{3.14 (1.68)}                 & \multicolumn{1}{l|}{3.68 (2.08)}    &  \\ \cline{1-4}
		&                                                  &                                                  &                                     & 
	\end{tabular}
	\caption{The average entry time in seconds and number of clicks to achieve partial and total correctness in the \textit{image-selection} scheme. The numbers in brackets are standard deviation.}
	\label{tab:threshold}
\end{table}

Furthermore, as an effort to improve the usability, we investigated another daily condition case involving four subjects (two females, two with glasses).
Each of them wore the camera in two consecutive days.
Then, they tried to solve the \textit{image-selection} authentication challenges whenever they wanted.
If we consider the password length $2-8$ images with $1-7$ valid images, the mean entry time is 4.67 seconds (standard deviation 3.17) and the mean number of clicks is 2.92 (standard deviation 1.89).
Nevertheless, in this case, we studied only the eight-image passwords because they had a balanced ratio of valid and invalid images.
When the users answered the challenges, we quantified the correctness of the current solution based on the number and position of chosen valid photos.
Table~\ref{tab:threshold} shows the user effort in terms of the average entry time (seconds) and number of clicks to achieve answers 50\%, 75\%, and 100\% of similarity comparing with the correct image-based authentication challenges.
Based on the observation, the users can be allowed to configure our system as a trade-off between security and usability, e.g. an input can be considered as \textit{valid} if it has 75\% of similarity with the correct passwords.
We consider this technique as an approach to reduce the entry time and the number of clicks.

\section{Threat model}
In contrast to other image-based authentication schemes where images are selected from a \textit{fixed} collection~\cite{2000_Dhamija_ImageAuthentication}~\cite{2009_Everitt_Passfaces}, our photo collection comes from ever-changing egocentric videos.
Therefore, our approach is resistant to smudge attacks and shoulder surfing. 
It is further not possible to steal the password from the user (as it is, for instance, with biometric-based schemes), since the authentication challenge is never the same and, in particular, the user is not aware of the authentication challenge ahead of time. 
This {\em enrollment phase} proceeds implicitly without the user's intervention.
Thus, generally speaking, the entropy of the graphical password space depends on the user's personal activities, i.e. contextual information.
If few different activities are conducted, the complexity of the authentication challenges generated degenerates as variation is low.
In such case (when very similar image features are observed), we suggest that either the history considered for password generation is extended, or fallback authentication is used.
This is useful also to reduce frustration for the legitimate user as we have observed in our experiments a tendency that low-variance behaviour might occasionally result in difficult to solve authentication challenges in which the exact order of similar and spatially close activities is hard to recall. 

A natural threat for our authentication scheme is an adversary following the user, remembering visual context, and bringing the locked device in her possession.
To successfully answer the image-based passwords, she or he is then required to know thoroughly the user's actions and positions.
While this is possible, we remark that it is difficult to achieve undetected, especially for a stranger, because the user may spend time in a personal space (e.g. a private office).
We remark that this issue has been addressed also by Shiraga~\textit{et al.} in~\cite{2012_Shiraga_CameraAuthentication}.
The authors deployed two cameras on a back-pack to capture what happens behind the wearer when walking.
Furthermore, even if the adversary obtains knowledge on the daily routine of the user, it is still challenging to guess the generated passwords without tracking the user suspiciously.
We have investigated these cases in a user study (see the box below). 

Another case is that the adversary has both the eyeglasses and the locked device in her possession for a duration which is long enough to generate an authentication challenge.
In this case, the image-generated authentication challenges would pose no protection. 
However, we suggest that the protocol is implemented together with a framework to verify the eyeglasses wearer (cf. figure~\ref{figureAuthenticationScheme}), so that the different user would be detected.
We therefore believe that this threat is easily mitigated for our system.

\noindent
\fbox{\begin{minipage}{.95\columnwidth}
\underline{\textit{Informed attackers: }}
In the object-interaction scenario (see Section~\ref{sec:room}), we conducted an attack in which the adversaries (2 attackers) had good knowledge on the environment (including locations of the furniture and suggested activities) and tried to solve others' authentication challenges.
These attackers reported that they leveraged two strategies to obtain the correct temporal order of images: (1) leveraging their knowledge on the suggested activities and the furniture arrangement to form a candidate image order, and (2) fixing an arrangement for every graphical password without paying attention to the images.
If they could not determine any answer, they tended to choose a random one.
The mean time spended on a single challenge was 54.91 seconds (standard deviation 67.04) with average 10.72 attempts (standard deviation 10.91) for utilizing random patterns.
In case of the second strategy, each image-based password took the attacker 64 seconds (standard deviation 56.12) with 22.36 attempts (standard deviation 20.82) on average.
Even though occasionally the attackers randomly selected the correct order, their effort was much higher than that of a legitimate user.
Hence, thresholds can be set on the user's effort (e.g. limits on entry time and number of attempts) to issue other authentication schemes (e.g. security questions) or lock the personal devices permanently.
      \end{minipage}
}

\section{Conclusion}

This paper presents a novel user authentication mechanism which takes advantage of egocentric videos recorded by wearable cameras.
We explain how to select video frames that are possible to discriminate in the timeline.
Those chosen images contain visual details that are meaningful to the users.
Two visual descriptors, CENTRIST~\cite{2011_Wu_CENTRIST} and PHOG~\cite{2007_Bosch_PHOG}, are aggregated to characterize an photo.
We explain the selection criteria and implement the algorithm which is based on image segmentation and clustering techniques.
They are organized into two graphical password schemes in different time scales: \textit{image-arrangement} and \textit{image-selection}.
In order to authenticate in the first scheme, an image sequence must be arranged into their chronological order.
In this case, the chosen images belong to short and instant activities rather than daily routines.
The second scheme leverages human memory on when specific events occur.
To log-in, users are required to identify which scenes have not appeared or which activities have not happened at a certain moment.
To evaluate our proposed approach, the user study is performed in three settings, including daily activities, object-interaction actions, and traveling scenarios:
\begin{itemize}
	\item The \textit{image-arrangment} scheme has the average entry time of 9.77 seconds in the object-interaction scenario and 9.79 seconds in two-day experiments
	\item The \textit{image-selection} scheme achieves the average entry time of 3.10 seconds in a three-week study and 4.67 seconds in two-day experiments
\end{itemize}
The experimental results show that the temporary and context-based graphical passwords are usable and compatible with touch-based interfaces.

\bibliographystyle{IEEEtran}

\end{document}